\documentclass[twocolumn]{article}

\usepackage{amsmath, amssymb, graphicx,natbib}
\usepackage{mathptmx}
\newcommand{\ti}[1]{^{\left(#1\right)}}
\newcommand{\lti}[2]{\mathbf{#1}^{\left(#2\right)}}
\newcommand{\gti}[2]{\boldsymbol{#1}^{\left(#2\right)}}
\newcommand{\tti}[2]{\bigl(\mathbf{#1}^{\left(#2\right)}\bigr)^{T}}
\newcommand{\gtti}[2]{\bigl(\boldsymbol{#1}^{\left(#2\right)}\bigr)^{T}}
\newcommand{\nn}{\nonumber\\}

\begin{document}
\title{Resonant forcing of select degrees of freedom of multidimensional chaotic map dynamics}
\author{Vadas Gintautas, Glenn Foster, and Alfred W. H\"{u}bler\\Center for Complex Systems Research,\\ Department of Physics, University of Illinois at Urbana-Champaign,\\ Urbana, Illinois 61801, USA\\
Email: vgintau2@uiuc.edu, gfoster@uiuc.edu, a-hubler@uiuc.edu}
\date{\today}
\maketitle
\begin{abstract}
We study resonances of multidimensional chaotic map dynamics.  We use the calculus of variations to determine the additive forcing function that induces the largest response, that is, the greatest deviation from the unperturbed dynamics.  We include the additional constraint that only select degrees of freedom be forced, corresponding to a very general class of problems in which not all of the degrees of freedom in an experimental system are accessible to forcing.  We find that certain Lagrange multipliers take on a fundamental physical role as the efficiency of the forcing function and the effective forcing experienced by the degrees of freedom which are not forced directly.  Furthermore, we find that the product of the displacement of nearby trajectories and the effective total forcing function is a conserved quantity.  We demonstrate the efficacy of this methodology with several examples.
\end{abstract}
\section{Introduction}
\label{sec:intro}
Sinusoidally driven nonlinear oscillators have been widely studied in contexts ranging from synchronization~\cite{eisenhammer90} and nonlinear response phenomena~\cite{morton70,siddiqi05} to stochastic resonance~\cite{kapral93,bulsara05} and nonlinear transport phenomena~\cite{wimberger05}.  Somewhat less studied are system identification via resonance curves of nonlinear systems~\cite{krempl92} and periodically driven chaotic systems~\cite{ruelle86}.  An area that has received much less attention is resonance phenomena of nonlinear systems due to aperiodic and chaotic forcing functions~\cite{mallick05}.  Plapp and H\"{u}bler~\cite{plapp90} and others~\cite{wargitsch95b} have used the calculus of variations to show that a special class of aperiodic driving forces can achieve a large energy transfer to a nonlinear oscillator.  Such nonsinusoidal resonant forcing functions yield a high signal-to-noise ratio which can be used for high-resolution system identification~\cite{chang91}.  In a recent paper, Foster, H\"{u}bler, and Dahmen~\cite{foster07} explored resonant forcing of chaotic map dynamics in which every degree of freedom in a multidimensional system is forced.

In this paper, we present a methodology for determining the resonant forcing of a multidimensional chaotic map in which only select degrees of freedom are forced.  This is motivated by the difficulty or impossibility of forcing all of the degrees of freedom in certain experiments.  For example, consider the forced one dimensional H\'{e}non map with a delay: $x^{\left(n+1\right)}=1-a\bigl(x^{\left(n\right)}\bigr)^{2}+cbx^{\left(n-1\right)}+F^{\left(n\right)}$.  This can be written as a two dimensional system with no delay, but then only one of the two dimensions is forced, corresponding to the presence of an $F^{\left(n\right)}$ term but the absence of an $F^{\left(n-1\right)}$ term.  Therefore, the method we present may be applied to a very general class of problems.  In practice, the calculations necessary to determine the optimal forcing are simpler than in the case where all degrees of freedom are forced, particularly if only one degree of freedom in a high dimension system is forced.  We show analytically that the resonant forcing functions are closely related to the unperturbed dynamics of the system in that the product of the displacement of nearby trajectories and the effective total forcing function is a conserved quantity.  Furthermore, we find that certain Lagrange multipliers take on a fundamental physical role as the efficiency of the forcing function and the effective forcing experienced by the degrees of freedom which are not forced directly.  We demonstrate the efficacy of the methodology with several examples.

\section{Resonant forcing of select degrees of freedom}
\label{sec:optforce}
We begin with the iterated map dynamics
\begin{equation}
\lti{x}{n+1}=\mathbf{f}\bigl(\lti{x}{n}\bigr)+\lti{F}{n},
\label{eq:xdyn}
\end{equation}
where $\mathbf{x}^{\left(n\right)}\in \mathbb{R}^{d}$ denotes the state of the $d$-dimensional system at the $n$th time step, with $n=0,1,\ldots,N-1$ and $\mathbf{F}^{\left(n\right)}\in \mathbb{R}^{d}$ denotes the forcing function at time step $n$.  This system has $d$ degrees of freedom.   We define the total forcing effort to be the magnitude of $\mathbf{F}$:
\begin{equation}
F^{2}=\sum^{N-1}_{n=0}{\bigl(\mathbf{F}^{\left(n\right)}\bigr)^{2}}.
\label{eq:fmagdefn}
\end{equation}
Given the corresponding unperturbed system $\lti{y}{n+1}=\mathbf{f}\bigl(\lti{y}{n}\bigr)$ with $\lti{y}{0}=\lti{x}{0}$, we define the final response as the deviation from the unperturbed dynamics
\begin{equation}
R^{2}\equiv\bigl(\lti{x}{N}-\lti{y}{N}\bigr)^{2}.
\label{eq:Rdefn}
\end{equation}
We require that $0\leq d_{u}<d$ degrees of freedom be unforced.  Without loss of generality, we choose to order the variables so that $x_{1},\ldots,x_{d_{u}}$ are unforced and $x_{d_{u}+1},\ldots,x_{d}$ are forced.  Thus we will require that
\begin{alignat}{2}
&F_{i}^{\left(n\right)}=0,&\qquad&\text{for $i=1,\ldots,d_{u}$ and $n=0,1,\ldots,N-1$.}
\label{eq:fzero}
\end{alignat}
The Lagrange function $L$ used to determine the forcing function that produces the largest response $R$ is
\begin{equation}
\begin{split}L=\frac{R^{2}}{2}+&\sum^{N-1}_{n=0}\Biggl\{\gti{\mu}{n}\Bigl[\lti{x}{n+1}-\mathbf{f}\bigl(\lti{x}{n}\bigr)-\lti{F}{n}\Bigr]\frac{}{}\Bigr.\\&-\Bigl.\frac{\lambda}{2}\Bigl[\bigl(\lti{F}{n}\bigr)^{2}-F^{2}\Bigr]-\lambda\sum^{d_{u}}_{j=1}\gamma\ti{n}_{j}F\ti{n}_{j}\Biggr\},\end{split}\label{eq:Ldefn}
\end{equation}
where $\lambda$, $\bigl\{\gti{\gamma}{n}_{1},\ldots,\gti{\gamma}{n}_{d_{u}}\bigr\}$, and $\bigl\{\mu\ti{n}_{1},\mu\ti{n}_{2},\ldots,\mu\ti{n}_{d}\bigr\}$ are Lagrange multipliers and $F$ is a constant.  We seek stationary points of $L$ corresponding to $\partial L/\partial x^{\left(n\right)}_{i}=0$ and $\partial L/\partial F^{\left(n\right)}_{i}=0$ for all $n$ and $i=1,\ldots,d$.  These equations of motion yield:
\begin{eqnarray}
\tti{J}{n+1}\gti{\mu}{n+1}-\gti{\mu}{n}&=&0\label{eq:mueqn}\\\lambda\lti{F}{n}+\gti{\mu}{n}+\lambda\gti{\Gamma}{n}&=&0,
\label{eq:secondFeqn}
\end{eqnarray}
where $J\ti{n}_{ij}=\bigl.\bigl(\partial f_{i}/\partial x_{j}\bigr)\bigr|_{\lti{x}{n}}$ is the Jacobi matrix evaluated at $\lti{x}{n}$.  We have also defined the vector $\boldsymbol{\Gamma}^{\left(n\right)}\equiv\sum^{d_{u}}_{j=1}\gamma_{j}^{\left(n\right)}\hat{\mathbf{e}}_{j}$, where $\hat{\mathbf{e}}_{j}$ is the unit basis vector in the direction of $x_{j}$.  The superscript $T$ indicates the transpose operator.  For $x^{\left(N\right)}$ we have the additional equation
\begin{equation}
\mathbf{x}^{\left(N\right)}-\mathbf{y}^{\left(N\right)}+\boldsymbol{\mu}^{\left(N-1\right)}=0.
\label{eq:mubc}
\end{equation}
We now define the quantity
\begin{equation} \mathbf{G}^{\left(n\right)}\equiv\mathbf{F}^{\left(n\right)}+\boldsymbol{\Gamma}^{\left(n\right)};\label{eq:Gdefn}
\end{equation}
after we eliminate the vector Lagrange multipliers $\bigl\{\gti{\mu}{0},\ldots,\gti{\mu}{N-1}\bigr\}$, the equations corresponding to the stationary points of Eq.~\eqref{eq:Ldefn} reduce to the same form as when all variables are forced [See \cite{foster07}]:
\begin{eqnarray}
\tti{J}{n+1}\lti{G}{n+1}&=&\lti{G}{n}\label{eq:GandJ}\\
\lti{x}{N}-\lti{y}{N}&=&\lambda\lti{G}{N-1}.\label{eq:GeqnNbc}
\end{eqnarray}
Accordingly, we identify $\mathbf{G}$ as the effective total forcing function; it reduces to the optimal forcing $\mathbf{F}$ when we remove the constraint in Eq.~\eqref{eq:fzero}.  We identify the Lagrange multipliers $\bigl\{\gti{\gamma}{n}_{1},\ldots\gti{\gamma}{n}_{d_{u}}\bigr\}$ to be the effective forcing experienced by the degrees of freedom $j$ for which $F\ti{n}_{j}=0$; this changes the trajectories of these degrees of freedom via the coupling in $\mathbf{f}$ rather than direct additive forcing via $\mathbf{F}$.  The control is stable if, on average, the displacement of nearby trajectories decreases.  Consider a trajectory given by Eq.~\eqref{eq:xdyn}, and a nearby trajectory given by $\lti{\tilde{x}}{n+1}=\mathbf{f}\bigl(\lti{\tilde{x}}{n}\bigr)+\lti{F}{n}$, where $\mathbf{x}$ and $\tilde{\mathbf{x}}$ are related by 
$\gti{\epsilon}{n}\equiv\lti{x}{n}-\lti{\tilde{x}}{n}$.  If we Taylor expand $\mathbf{f}\bigl(\mathbf{x}\bigr)$ for small $\boldsymbol{\epsilon}$, we obtain
\begin{equation}
\gti{\epsilon}{n+1}=\lti{J}{n}\gti{\epsilon}{n}.\label{eq:conserveddyn}
\end{equation}
Multiplying both sides of the transpose of Eq.~\eqref{eq:GandJ} by $\gti{\epsilon}{n}$, we have\\ $\tti{G}{n+1}\lti{J}{n+1}\gti{\epsilon}{n+1}=\tti{G}{n}\gti{\epsilon}{n+1}$.  Using Eq.~\eqref{eq:conserveddyn}, this becomes
\begin{equation}
\tti{G}{n+1}\gti{\epsilon}{n+2}=\tti{G}{n}\gti{\epsilon}{n+1},
\end{equation}
a quantity that is invariant for all $n$.  We define this to be the conserved quantity $P$:
\begin{equation}
\lti{G}{0}\cdot\gti{\epsilon}{1}\equiv P=\lti{G}{n}\cdot\gti{\epsilon}{n+1},
\label{eq:conserved}
\end{equation}
and note that $P$ depends on the observables $\mathbf{x}$ and $\mathbf{F}$ as well as the Lagrange multipliers $\boldsymbol{\Gamma}$, which we have identified as the effective indirect forcing of certain degrees of freedom.  This further reinforces the idea that $\mathbf{G}$ represents the effective forcing experienced by the system, taking into account the coupling via $\mathbf{f}$.  As in the case where all degrees of freedom are forced, $P$ is conserved even if the unperturbed dynamics is chaotic or periodic.

\subsection{Resonant forcing functions with small magnitude}
\label{sec:smallforce}
For weak forcing, we can iterate Eq.~\eqref{eq:xdyn} and Taylor expand for small $\mathbf{F}$.  We obtain [see Eq.~\eqref{eq:RMGproof} in the Appendix]
\begin{equation}
\lti{x}{N}-\lti{y}{N}=M\lti{G}{N-1}-\Omega,
\label{eq:xyMomegaG}
\end{equation}
where we have defined $M\equiv I+\sum_{n=1}^{N-1}J^{\left(N-1\right)}\cdots J^{\left(N-n\right)}\bigl(J^{\left(N-n\right)}\bigr)^{T}\cdots\bigl(J^{\left(N-1\right)}\bigr)^{T}$ where $I$ is the identity matrix and
\begin{equation}
\Omega\equiv\gti{\Gamma}{N-1}+\lti{J}{N-1}\gti{\Gamma}{N-2}+\cdots+\bigl(\lti{J}{N-1}\cdots\lti{J}{1}\bigr)\gti{\Gamma}{0}.\label{eq:Omegadefn}
\end{equation}
Using Eq.~\eqref{eq:GeqnNbc}, we obtain 
\begin{equation}
M\lti{G}{N-1}-\Omega=\lambda\mathbf{G}^{\left(N-1\right)}.
\label{eq:MomegaG}
\end{equation}
Eqs.~\eqref{eq:fmagdefn},~\eqref{eq:fzero},~\eqref{eq:MomegaG},~and~\eqref{eq:GandJ} form a complete system of equations whereby the unknown forcing amplitudes $\bigl\{F\ti{0}_{i},\ldots,F\ti{N-1}_{i}\bigr\}$ with $i=(d_{u}+1),\ldots,d$ and Lagrange multipliers $\lambda$ and $\bigl\{\gamma\ti{0}_{j},\ldots,\gamma\ti{N-1}_{j}\bigr\}$ with $j=1,\ldots,d_{u}$ can be uniquely determined.  At this point it is possible to write the final response in terms of the total effort:
\begin{equation}
R^{2}=\lambda F^{2}
\label{eq:RlambdaF};
\end{equation}
the details of the proof are in the Appendix [see Eq.~\eqref{eq:RlambdaFproof}].  Thus $\lambda$ is the effective efficiency $R^{2}/F^{2}$ of the forcing function.  One approach to solving this system of equations is to treat Eq.~\eqref{eq:MomegaG} as an inhomogeneous eigenvalue problem.  To find the optimal forcing, we first find the eigenvectors $\left\{\mathbf{v}_{1},\ldots,\mathbf{v}_{d}\right\}$ and eigenvalues $\lambda_{i}$ of the homogeneous problem $\left(M-\lambda I\right)\lti{G}{N-1}=0$.  Then we can build the inverse $H_{\lambda}\equiv\left(M-\lambda I\right)^{-1}$:
\begin{equation}
H_{\lambda}=\sum^{d}_{j=1}{\frac{\mathbf{v}_{j}\mathbf{v}_{j}^{T}}{\lambda_{j}-\lambda}},
\end{equation}
in which $\lambda$ is not yet determined.  Then the solutions to the inhomogeneous problem are
\begin{equation}
\mathbf{G}^{\left(N-1\right)}=H_{\lambda}\Omega.
\label{eq:GHOmega}
\end{equation}
Eq.~\eqref{eq:GandJ} can then be used to build a set of equations
\begin{equation}
\lti{G}{n}=\tti{J}{n+1}\cdots\tti{J}{N-1}H_{\lambda}\Omega,\qquad n=0,\ldots,N-2
\end{equation}
whereby $\lambda$, $\lti{F}{n}$ and $\gti{\Gamma}{n}$ can be determined for all $n$ without matrix inversion.  This method allows a natural connection to the case in which all degrees of freedom are forced, that is, $d_{u}\rightarrow 0$ in Eq.~\eqref{eq:fzero}.  For this case, $\Omega=0$ and Eq.~\eqref{eq:MomegaG} reduces to the homogeneous eigenvalue problem $\left(M-\lambda I\right)\lti{F}{N-1}=0$ with solutions $\lti{F}{N-1}=\left\{\mathbf{v}_{1},\ldots,\mathbf{v}_{d}\right\}$, each corresponding to an eigenvalue $\lambda_{i}$.  For the homogeneous case, if we set $\lambda=\max\left\{\lambda_{i}\right\}$ to be the largest eigenvalue of $M$ then Eq.~\eqref{eq:RlambdaF} holds and the Lagrange multiplier $\lambda$ assumes the same meaning as in the inhomogeneous case, namely, the effective efficiency of $\mathbf{F}$.

\section{Examples}
\subsection{Resonances of coupled shift maps}
\label{sec:shift}
\begin{figure}
	\centering	\includegraphics[width=0.85\linewidth]{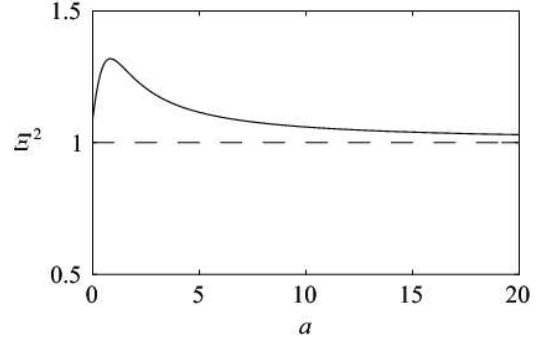}
	\caption{The ratio of final responses $\Xi^{2}$ versus the parameter $a$ for the same total effort $F=0.001$ and $k=0.3000$.  For $a>0$ and $k\neq 0$, the final response is greater when both degrees of freedom are forced (corresponding to $\Xi^{2}>1$).  Note that $\Xi^{2}\rightarrow 1$ as $a\rightarrow\infty$ or $k\rightarrow 0$.}
	\label{fig:methodratio}
\end{figure}

We consider the mapping function for coupled shift maps:
\begin{equation}
\left(
\begin{array}{c}
x^{\left(n+1\right)}_{1}\\
x^{\left(n+1\right)}_{2}\\
\end{array}
\right)=\left(
\begin{array}{c}
\text{mod}(a x^{\left(n\right)}_{1}+kx^{\left(n\right)}_{2})\\
\text{mod}(a x^{\left(n\right)}_{2}+kx^{\left(n\right)}_{1})\\
\end{array}
\right)+\left(
\begin{array}{c}
F^{\left(n\right)}_{1}\\
F^{\left(n\right)}_{2}\\
\end{array}\right)
\end{equation}
and require that only $F_{2}$ be forced, that is, $F^{\left(n\right)}_{1}=0$ for all $n$.  Accordingly, $\boldsymbol{\Gamma}^{\left(n\right)}=\gamma^{\left(n\right)}\hat{\mathbf{e}}_{1}$, where we have defined $\gamma\equiv\gamma_1$.  Since the Jacobi matrix
\begin{equation}
\mathbf{J}^{\left(n\right)}\equiv\mathbf{J}=\left(\begin{array}{cc}
a & k\\
k & a\\
\end{array}\right)=\mathbf{J}^{T}\label{eq:shiftj}
\end{equation}
is symmetric and constant, the eigenvectors of $M$ for any $N$ are $\mathbf{v_{\pm}}=\bigl(\begin{smallmatrix}\pm 1/\sqrt{2}\\1/\sqrt{2}\end{smallmatrix}\bigr)$.  We denote the two eigenvalues of $M$ by $\lambda{\pm}$, with $\lambda_{+}\ge\lambda_{-}$.  Because of the nature of the eigenvectors, we can always write $M$ explicitly:
\begin{equation}
M=\left(\begin{array}{cc}
\lambda^{+} & \lambda^{-}\\
\lambda^{-} & \lambda^{+}\\
\end{array}\right),
\label{eq:mlambda}
\end{equation}
where we have defined the quantities $\lambda^{\pm}\equiv\left(\lambda_{+}\pm\lambda_{-}\right)/2$.  It can be shown that for $N=2^{b}$ with $b\in\mathbb{Z^{+}}$, the eigenvalues of $M$ are $\lambda_{\pm}=\prod_{i=1}^{b}\Bigl[1+\bigl(a\pm k\bigr)^{2^{i}}\Bigr]$.  As an example, we set $N=2$ so that the eigenvalues of $M$ are $\lambda_{\pm}=1+(a\pm k)^{2}$.  Likewise, we find $\Omega=\bigl(\begin{smallmatrix}\gamma\ti{1}+a\gamma\ti{0}\\k\gamma\ti{0}\end{smallmatrix}\bigr)$ using Eq.~\eqref{eq:Omegadefn}.  From Eq.~\eqref{eq:fzero} we impose the additional constraints $F\ti{0}_{1}=F\ti{1}_{1}=0$.  If we define
\begin{equation}
\beta_{\pm}\equiv\pm\sqrt{(1+a^{2})^{2}+2k^{2}(a^{2}-1)+k^{4}},
\end{equation}
we can write the solutions to Eq.~\eqref{eq:MomegaG}:
\begin{align}
&\lambda=\frac{1}{2}\bigl(1+a^{2}+k^{2}-\beta_{\pm}\bigr),\label{eq:shiftlambda}\\
&\gamma\ti{0}=-\frac{1}{2k}\bigl(1+a^{2}-3k^{2}+\beta_{\pm}\bigr)F\ti{1}_{2},\label{eq:shiftgamma0}\\
&\gamma\ti{1}=-\frac{1}{2ak}\bigl(1+a^{2}-k^{2}+\beta_{\pm}\bigr)F\ti{1}_{2},\label{eq:shiftgamma1}\\
&F\ti{0}_{2}=-\frac{1}{2a}\bigl(1-a^{2}-k^{2}+\beta_{\pm}\bigr)F\ti{1}_{2}.\label{eq:shiftf20}
\end{align}
Using the normalization condition in Eq.~\eqref{eq:fmagdefn}, we can determine the magnitude of $F\ti{1}_{2}$:
\begin{equation}
F\ti{1}_{2}=\cfrac{2aF}{\sqrt{4a^{2}+\bigl(1-a^{2}-k^{2}+\beta_{\pm}\bigr)^{2}}}.
\label{eq:shiftf21}
\end{equation}
We use Eqs.~\eqref{eq:RlambdaF}~and~\eqref{eq:shiftlambda} to find the final response
\begin{equation}
\frac{R^{2}}{F^{2}}=\lambda=\frac{1}{2}\bigl(1+a^{2}+k^{2}-\beta_{\pm}\bigr).
\label{eq:shiftR}
\end{equation}
From this equation it can be shown that $\beta_{+}$ gives the minimum response while $\beta_{-}$ gives the maximum response.  For the special case $k\rightarrow 0$, corresponding to the case of two uncoupled shift maps where one is forced but the other is not, the response reduces to the simple form $R^{2}/F^{2}\rightarrow 1+a^{2}$.  We may also compare the effectiveness of only forcing $x_{2}$ to the effectiveness of forcing both $x_{1}$ and $x_{2}$.  This corresponds to removing the constraint in Eq.~\eqref{eq:fzero} and having $\gamma\ti{n}_{j}=0$ for all $j$ and $n$ in Eq.~\eqref{eq:Ldefn}.  We will mark solutions of this system with a ($\sim$) to avoid confusion with the selectively forced system.  Thus $\mathbf{\tilde{G}}=\mathbf{\tilde{F}}$ and Eq.~\eqref{eq:MomegaG} reduces to $M\lti{\tilde{F}}{N-1}=\lambda\mathbf{\tilde{F}}^{\left(N-1\right)}$.  We solve this system to find $\tilde{\lambda}$ as well as the individual components of $\mathbf{\tilde{F}}$ for each timestep with $N=2$:
\begin{align}
&\tilde{\lambda}=1+\bigl(a\pm k\bigr)^{2},\\
&\tilde{F}\ti{0}_{1}=-\bigl(a\pm k\bigr)^{2}F\ti{1}_{2},\\
&\tilde{F}\ti{1}_{1}=\pm F\ti{1}_{2},\\
&\tilde{F}\ti{0}_{2}=\bigl(a\pm k\bigr)^{2}F\ti{1}_{2},\\
&\tilde{F}\ti{1}_{2}=\frac{F}{\sqrt{2+2\bigl(a\pm k\bigr)^{2}}}.
\end{align}
Note that $\tilde{F}\ti{n}_{1}\neq \gamma\ti{n}$; the effective indirect forcing of $x_{1}$ when only $x_{2}$ is forced is not simply equal to the optimal direct forcing of $x_{1}$ when both $x_{1}$ and $x_{2}$ are forced.  Using Eq.~\eqref{eq:RlambdaF}, we can calculate the final response:
\begin{equation}
\frac{\tilde{R}^{2}}{F^{2}}=\tilde{\lambda}=1+\bigl(a\pm k\bigr)^{2},
\label{eq:fosterR}
\end{equation}
where as before, the $(+)$ sign in front of $k$ corresponds to a maximum final response for $a>0$, while a $(-)$ in the same position corresponds to a minimum final response.  We will assume $a>0$ and use $(+)$ henceforth for these results.  This quantity also reduces to $1+a^{2}$ when $k\rightarrow 0$.  To compare the effectiveness of forcing only one degree of freedom to forcing both, we assume that the total forcing effort $F$ is the same in both cases, then use Eqs.~\eqref{eq:shiftR}~and~\eqref{eq:fosterR} to obtain the ratio of final responses:
\begin{equation}
\Xi^{2}\equiv\frac{\tilde{R}^{2}}{R^{2}}=\frac{\tilde{\lambda}}{\lambda}=\frac{2\bigl[1+\bigl(a+k\bigr)^{2}\bigr]}{1+a^{2}+k^{2}+\beta_{-}}.
\label{eq:methodratio}
\end{equation}
Note that $\Xi^{2}\rightarrow 1$ as $k\rightarrow 0$.  We plot Eq.~\eqref{eq:methodratio} in Fig.~\ref{fig:methodratio} for $k=0.3000$.  Notice that $\Xi^{2}>1$ for all $a$; that is, for the same total effort $F$, there will always be a greater final response if all the degrees of freedom are forced.  This is reasonable, given that when both $x_{1}$ and $x_{2}$ are forced there is a more uniform flow of energy into the system than when only $x_{2}$ is forced and $x_{1}$ is passive.  However, $\Xi^{2}\propto 1$ for small $a$ and $\Xi^{2}\rightarrow 1$ for large $a$.  Since $\tilde{R}^{2}$ is on the same order of magnitude as $R^{2}$, for certain experimental situations it may be sufficient (and presumably easier) to build an apparatus in which only one degree of freedom is forced rather than all.  For $a<0$, we choose the $(-)$ sign before $k$ in Eq.~\eqref{eq:fosterR}.  The result is the same; $\Xi^{2}>1$ for small negative $a$ but $\Xi^{2}\rightarrow 1$ for $a\rightarrow -\infty$.

\begin{figure}
	\centering	\includegraphics[width=0.85\linewidth]{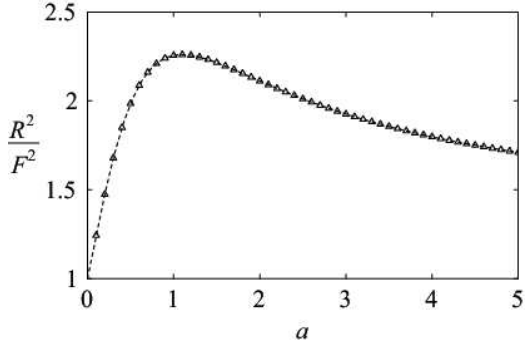}
	\caption{Shift map resonance curve for final response $R^{2}/F^{2}$ versus the parameter $a$.  The total effort is $F=0.001$ and the other control parameters are $a_{0}=0.6000$ and $k=0.3000$.  The maximum of the curve is at $a=a_{0}$ within machine precision.  The solid line indicates the analytical expression given in Eq.~\eqref{eq:shiftRa0} and the triangles indicate the results of a numerical simulation with $x\ti{0}_{1}=x\ti{0}_{2}=0.1000$ as the initial condition.}
	\label{fig:shiftres}
\end{figure}

Furthermore, for this system we can explicitly verify Eq.~\eqref{eq:conserved}: $\tti{G}{1}\gti{\epsilon}{2}=\tti{G}{0}\gti{\epsilon}{1}=P$.
Using Eq.~\eqref{eq:conserveddyn}, we can write this as
\begin{equation}
\tti{G}{1}\lti{J}{1}\gti{\epsilon}{1}=\tti{G}{0}\gti{\epsilon}{1}.
\label{eq:shiftconserved}
\end{equation}
Using Eqs.~\eqref{eq:shiftgamma1}~and~\eqref{eq:shiftj}, for the left hand side of Eq.~\eqref{eq:shiftconserved} we obtain
\begin{align}
\tti{G}{1}\lti{J}{1}\gti{\epsilon}{1}&=\frac{-F\ti{1}_{2}}{2ak}\bigl[a\bigl(1+a^{2}-3 k^{2}+\beta_{\pm}\bigr)\epsilon\ti{1}_{1}\nn
&\qquad+k\bigl(1-a^{2}- k^{2}+\beta_{\pm}\bigr)\epsilon\ti{1}_{2}\bigr]\equiv P.\label{eq:GPleft}
\end{align}
Using Eqs.~\eqref{eq:shiftgamma0}~and~\eqref{eq:shiftf20}, for the right hand side we find
\begin{align}
\tti{G}{0}\gti{\epsilon}{1}&=\frac{-F\ti{1}_{2}}{2ak}\bigl[a\bigl(1+a^{2}-3 k^{2}+\beta_{\pm}\bigr)\epsilon\ti{1}_{1}\nn
&\qquad+k\bigl(1-a^{2}- k^{2}+\beta_{\pm}\bigr)\epsilon\ti{1}_{2}\bigr]=P.\label{eq:GPright}
\end{align}
Thus for the coupled shift maps with $N=2$ we are able to analytically verify that $P$ is a conserved quantity.  Now consider a coupled shift map with $a=a_{0}$ driven by a forcing function that is described by Eqs.~\eqref{eq:shiftf20} and \eqref{eq:shiftf21}.  The response curve as a function of $a$ can by found by starting with Eq.~\eqref{eq:Rdefn} and iterating Eq.~\eqref{eq:xdyn}:
\begin{equation}
R^{2}=k^{2}\bigl[F\ti{0}_{2}\bigr]^{2}+\bigl[a_{0}F\ti{0}_{2}+F\ti{1}_{2}\bigr]^{2},
\label{eq:shiftRa0}
\end{equation}
where $F\ti{0}_{2}$ and $F\ti{1}_{2}$ are functions of $a$ as given by Eqs.~\eqref{eq:shiftf20}~and~\eqref{eq:shiftf21}, respectively.  We plot Eq.~\eqref{eq:shiftRa0} as a function of $a$ in Fig.~\ref{fig:shiftres}.  In the same figure we plot the results of a numerical simulation in which the final response is found by using $F\ti{0}_{2}$ and $F\ti{1}_{2}$ as the forcing functions and iterating the system for two time steps.  The initial condition used is $x\ti{0}_{1}=x\ti{0}_{2}=0.1000$ and the total effort is $F=0.001$. 

\subsection{Resonances of the one dimensional H\'{e}non map with delay}
\label{sec:henon}
The forced H\'{e}non map with delay $x^{\left(n+1\right)}=1-a\bigl(x^{\left(n\right)}\bigr)^{2}+cbx^{\left(n-1\right)}+F^{\left(n\right)}$ can be written as the equivalent two-dimensional system
\begin{equation}\left(\begin{array}{c}
x^{\left(n+1\right)}_{1}\\
x^{\left(n+1\right)}_{2}\\
\end{array}\right)=\left(\begin{array}{c}
bx^{\left(n\right)}_{2}\\
1-a\bigl(x^{\left(n\right)}_{2}\bigr)^{2}+cx^{\left(n\right)}_{1}\\
\end{array}\right)+\left(\begin{array}{c}
F^{\left(n\right)}_{1}\\
F^{\left(n\right)}_{2}\\
\end{array}\right)\label{eq:henondyn}
\end{equation}
and require that only $F_{2}$ be forced, that is, $F^{\left(n\right)}_{1}=0$ for all $n$.  Accordingly, $\boldsymbol{\Gamma}^{\left(n\right)}=\gamma^{\left(n\right)}\hat{\mathbf{e}}_{1}$, where we have defined $\gamma\equiv\gamma_1$.  The Jacobi matrix is 
\begin{equation}
\mathbf{J}^{\left(n\right)}=\left(\begin{array}{cc}
0 & b \\
c & -2ax^{\left(n\right)}_{2}\\
\end{array}\right);
\end{equation}
while for $N=2$ the matrix $M$ is given by
\begin{eqnarray}
M&=&\left(\begin{array}{cc}
1+b^{2} & -2abx^{\left(1\right)}_{2}\\
-2abx^{\left(2\right)}_{2} & 1+c^{2}+4a^{2}\bigl(x^{\left(1\right)}_{2}\bigr)^{2}\\
\end{array}\right),\label{eq:henonM}
\end{eqnarray}
where $x^{\left(1\right)}_{2}\approx y\ti{1}_{2}=1+cx^{\left(0\right)}_{1}-a\bigl[x^{\left(0\right)}_{2}\bigr]^{2}$.  Using Eq.~\eqref{eq:Omegadefn}, we find $\Omega=\bigl(\begin{smallmatrix}\gamma\ti{1}\\c\gamma\ti{0}\end{smallmatrix}\bigr)$.  We now define 
\begin{align}
&\alpha\equiv 2a\Bigl[1+cx\ti{0}_{1}-a \bigl(x\ti{0}_{2}\bigr)^{2}\Bigr],\\
&\beta_{\pm}\equiv\pm\sqrt{b^{4}+2 b^{2}\bigl(\alpha^{2}-1\bigr)+\bigl(1+\alpha^{2}\bigr)^{2}}.
\end{align}
As with the coupled shift maps, solving Eqs.~\eqref{eq:GandJ} and \eqref{eq:MomegaG}, and \eqref{eq:fzero} simultaneously yields the following:
\begin{align}
&\lambda=\frac{1}{2}\bigl(1+b^{2}+\alpha^{2}+\beta_{\pm}\bigr),\label{eq:henonlambda}\\
&\gamma\ti{0}=cF\ti{1}_{2},\\
&\gamma\ti{1}=\frac{1}{2b\alpha}\bigl(1-b^{2}+\alpha^{2}-\beta_{\pm}\bigr)F\ti{1}_{2},\\
&F\ti{0}_{2}=\frac{1}{2\alpha}\bigl(1-b^{2}-\alpha^{2}-\beta_{\pm}\bigr)F\ti{1}_{2}.\label{eq:henonf20}
\end{align}
Using the normalization condition in Eq.~\eqref{eq:fmagdefn}, we can determine the magnitude of $F\ti{1}_{2}$:
\begin{equation}
F\ti{1}_{2}=\frac{2\alpha F}{\sqrt{4\alpha^{2}+\bigl(1-b^{2}-\alpha^{2}-\beta_{-}\bigr)^{2}}}
\label{eq:henonf21}
\end{equation}

\begin{figure}
	\centering	\includegraphics[width=0.85\linewidth]{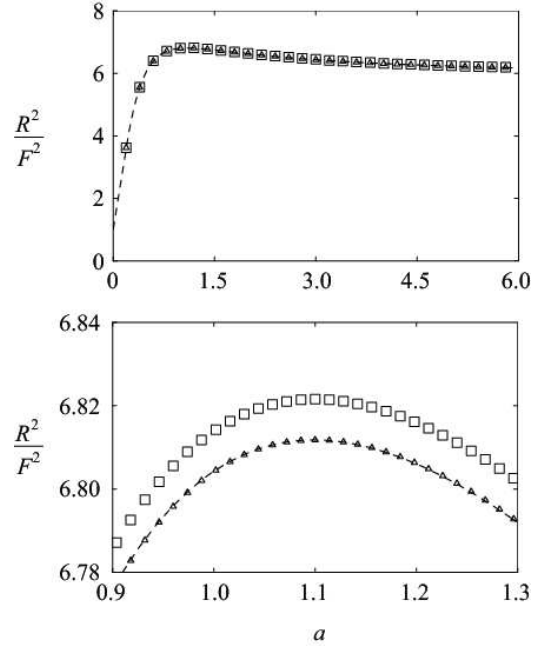}
	\caption{H\'{e}non map resonance curve showing final response $R^{2}/F^{2}$ versus the parameter $a$.  The total effort is $F=0.001$, the other control parameters are $a_{0}=1.1000$ and $k=0.3000$, and the initial conditions are $x\ti{0}_{1}=x\ti{0}_{2}=0.1000$.  The solid line indicates the analytical expression given in Eq.~\eqref{eq:henonRa0} and the triangles indicate the results of a numerical simulation of the simplified system in which the Jacobi matrix of the driven system is approximated by the Jacobi matrix of the unperturbed system.  The boxes indicate a numerical simulation of the exact system.  The approximate analytical result and the approximate numerical simulation curves have a maximum at $a=1.0991$.  The exact numerical simulation has a maximum at $a=1.1000$, which is the expected peak for this system.  (a) shows the full range of the experiment, $0\leq a\leq 6.0$, while (b) shows in detail the region near the maximum for all three curves.  (b) shows that the exact result has a greater maximum response than the approximate results, as expected.}
	\label{fig:henonboth}
\end{figure}

We can also use Eqs.~\eqref{eq:RlambdaF}~and~\eqref{eq:henonlambda} to find the final response
\begin{equation}
\frac{R^{2}}{F^{2}}=\lambda=\frac{1}{2}\bigl(1+b^{2}+\alpha^{2}+\beta_{\pm}\bigr)
\end{equation}
From this equation it can be shown that $\beta_{+}$ gives the maximum response while $\beta_{-}$ gives the minimum response provided $a>0$.  Now consider, as with the coupled shift map, a H\'{e}non map described by Eq.~\eqref{eq:henondyn}, only with $a\rightarrow a_{0}$.  Then we can find the final response as a function of $F\ti{0}_{2}$ and $F\ti{1}_{2}$
by starting with Eq.~\eqref{eq:Rdefn} and iterating Eq.~\eqref{eq:xdyn}:
\begin{equation}
R^{2}=b^{2}\bigl[F\ti{0}_{2}\bigr]^{2}+\Bigl\{F\ti{1}_{2}-a_{0}F\ti{0}_{2}\Bigl[F\ti{0}_{2}+2+2cx\ti{0}_{1}-2a_{0}\bigl(x\ti{0}_{2}\bigr)^{2}\Bigr]\Bigr\}^{2},
\label{eq:henonRa0}
\end{equation}
where $F\ti{0}_{2}$ and $F\ti{1}_{2}$ are functions of $a$ as given by Eqs.~\eqref{eq:henonf20}~and~\eqref{eq:henonf21}, respectively.  We plot Eq.~\eqref{eq:henonRa0} as a function of $a$ in Fig.~\ref{fig:henonboth}.  Since we use the approximation $x\ti{1}_{2}\approx y\ti{1}_{2}$ in Eq.~\eqref{eq:henonM}, we are able to solve Eqs.~\eqref{eq:GandJ},~\eqref{eq:MomegaG},~and~\eqref{eq:fzero} analytically.  With this approximation and with $a_{0}=1.1000$, the $R^{2}$ has a maximum at $a=1.0991$.  If instead we substitute the exact expression $x^{\left(1\right)}_{2}=1-a\bigl[x^{\left(0\right)}_{2}\bigr]^{2}+cx^{\left(0\right)}_{1}+F\ti{0}_{2}$ into Eq.~\eqref{eq:henonM}, it is possible to solve Eqs.~\eqref{eq:GandJ},~\eqref{eq:MomegaG},~and~\eqref{eq:fzero} numerically for $F\ti{n}_{2}$.  Then we recover the expected maximum at $a=1.1000$.  Furthermore, we find that for $F=0.001$, at maximum the numerical solution of the exact system gives $R^{2}/F^{2}=6.8215$, while the analytical approximate solution gives $R^{2}/F^{2}=6.8118$.  This is consistent with the claim that the methodology presented in this work will give the maximum response for a fixed effort.  Since any forcing other than the exact solution should be suboptimal, we expect a smaller final response when any approximations are used.

\section{Conclusion}
\label{sec:conclusion}
We study resonances of forced multidimensional chaotic map dynamics.  We constrain the total forcing effort to be fixed [see Eq.~\eqref{eq:fmagdefn}] and seek the forcing function which produces the largest response [Eq.~\eqref{eq:Rdefn}], subject to the additional constraint that certain degrees of freedom are not directly forced [Eq.~\eqref{eq:fzero}].  To determine this forcing function, we seek the stationary points of the Lagrange function [Eq.~\eqref{eq:Ldefn}] and thereby obtain equations which determine the dynamics of the forcing function [see Eqs.~\eqref{eq:GandJ}~and~\eqref{eq:GeqnNbc}].  From these equations we identify the effective total forcing to be a vector comprising the direct forcing and the Lagrange multipliers that represent the effective indirect forcing of certain degrees of freedom [see Eq.~\eqref{eq:Gdefn}].  We demonstrate that the product of the effective forcing and the displacement of nearby trajectories is a conserved quantity [Eq.~\eqref{eq:conserved}].  In the case of small forcing, we show that another Lagrange multiplier represents the efficiency of the forcing function [see Eq.~\eqref{eq:RlambdaF}].  In the limit that we set the number of unforced degrees of freedom to be zero, all of the results reduce to the homogeneous case.  The methodology presented can be applied to a very general class of problems in which not all of the degrees of freedom in an experimental system are accessible to forcing.  

We demonstrate the effectiveness of the methodology with several examples.  We compare forcing one degree of freedom in a system of two coupled shift maps to forcing both degrees and show that the final response is greater but on the same order of magnitude when both are forced [see Fig.~\ref{fig:methodratio}].  We present a resonance curve for the coupled shift map in Fig.~\ref{fig:shiftres} and verify explicitly that the optimal effective forcing complements the separation of nearby trajectories [see Eqs.~\eqref{eq:GPleft}~and~\eqref{eq:GPright}].  We also apply this method to a forced one dimensional H\'{e}non map with a delay [Eq.~\eqref{eq:henondyn}], a problem which cannot be solved using the homogeneous case in which all degrees of freedom are forced.  We solve for the optimal forcing function analytically for two time steps by approximating the Jacobi matrix of the forced system by the Jacobi matrix of the corresponding unperturbed system.  We also solve this system numerically without approximations and demonstrate that the exact solution reproduces the correct peak in the resonance curve and has a greater final response at maximum [see Fig.~\ref{fig:henonboth}].  Thus we show that the method may be used for system identification.  In the future we plan to compare the effectiveness of this methodology for system identification to that of other methods such as periodic driving~\cite{ruelle86} and coupling a test system to a virtual model with tunable parameters~\cite{gintautas07}.

\begin{section}{Acknowledgements}
\label{sec:ack}
The authors thank U.H. Gerlach for useful discussions concerning inhomogeneous eigenvalue problems and S. Raymond for other helpful input.  This work was supported by the National Science Foundation Grant Nos. NSF PHY 01-40179, NSF DMS 03-25939 ITR, and NSF DGE 03-38215.
\end{section}

\onecolumn
\section{Appendix: Proofs}
\label{sec:proofs}
Proof of Eq.~\eqref{eq:xyMomegaG}:  Using Eq.~\eqref{eq:GandJ}~and~\eqref{eq:Gdefn},
\begin{align}
\lti{F}{N-n}&=\tti{J}{N-n+1}\bigl[\lti{F}{N-n+1}+\gti{\Gamma}{N-n+1}\bigr]-\gti{\Gamma}{N-n}\nn
&=\tti{J}{N-n+1}\Bigl\{\tti{J}{N-n+2}\bigl[\lti{F}{N-n+2}+\gti{\Gamma}{N-n+2}\bigr]\gti{\Gamma}{N-n+1}\Bigr\}\nn
&\qquad+\tti{J}{N-n+1}\gti{\Gamma}{N-n+1}-\gti{\Gamma}{N-n}\nn
&=\tti{J}{N-n+1}\tti{J}{N-n+2}\lti{F}{N-n+2}\nn
&\qquad+\tti{J}{N-n+1}\tti{J}{N-n+2}\gti{\Gamma}{N-n+2}\nn
&\qquad-\tti{J}{N-N+1}\gti{\Gamma}{N-n+1}+\tti{J}{N-n+1}\gti{\Gamma}{N-n+1}-\gti{\Gamma}{N-n}\nn
&=\tti{J}{N-n+1}\tti{J}{N-n+2}\bigl[\lti{F}{N-n+2}+\gti{\Gamma}{N-n+2}\bigr]-\gti{\Gamma}{N-n}\nn
&\cdots\nn
\lti{F}{N-n}&=\tti{J}{N-n+1}\cdots\tti{J}{N-1}\bigl[\lti{F}{N-1}+\gti{\Gamma}{N-1}\bigr]-\gti{\Gamma}{N-n}
\end{align}

Using this result,
\begin{align}
\lti{x}{N}-\lti{y}{N}&=\lti{F}{N-1}+\lti{J}{N-1}\lti{F}{N-2}+\lti{J}{N-1}\lti{J}{N-2}\lti{F}{N-3}\nn
&\qquad+\cdots+\lti{J}{N-1}\cdots\lti{J}{1}\lti{F}{0}\nn
&=\bigl[\lti{F}{N-1}+\gti{\Gamma}{N-1}\bigr]+\lti{J}{N-1}\tti{J}{N-1}\bigl[\lti{F}{N-1}+\gti{\Gamma}{N-1}\bigr]+\cdots\nn
&\qquad+\lti{J}{N-1}\cdots\lti{J}{1}\tti{J}{1}\cdots\tti{J}{N-1}\bigl[\lti{F}{N-1}+\gti{\Gamma}{N-1}\bigr]\nn
&\qquad-\Bigl\{\gti{\Gamma}{N-1}+\lti{J}{N-1}\gti{\Gamma}{N-2}+\cdots+\bigl[\lti{J}{N-1}\cdots\lti{J}{1}\bigr]\gti{\Gamma}{0}\Bigr\}\nn
\lti{x}{N}-\lti{y}{N}&=M\lti{G}{N-1}-\Omega\label{eq:RMGproof}
\end{align}

Proof of Eq.~\eqref{eq:RlambdaF}:
\begin{align}
F^{2}&=\tti{F}{N-1}\lti{F}{N-1}+\tti{F}{N-2}\lti{F}{N-2}+\cdots+\tti{F}{0}\lti{F}{0}\nn
&=\Bigl[\tti{G}{N-1}-\gtti{\Gamma}{N-1}\Bigr]\Bigl[\lti{G}{N-1}-\gti{\Gamma}{N-1}\Bigr]\nn
&\qquad+\Bigl[\tti{G}{N-2}-\gtti{\Gamma}{N-2}\Bigr]\Bigl[\lti{G}{N-2}-\gti{\Gamma}{N-2}\Bigr]\nn
&\qquad+\cdots+\Bigl[\tti{G}{0}-\gtti{\Gamma}{0}\Bigr]\Bigl[\lti{G}{0}-\gti{\Gamma}{0}\Bigr]\nn
&=\tti{G}{N-1}\lti{G}{N-1}+\cdots+\tti{G}{0}\lti{G}{0}\nn
&\qquad-\tti{G}{N-1}\gti{\Gamma}{N-1}+\cdots+\tti{G}{0}\gti{\Gamma}{0}\nn
&\qquad-\gtti{\Gamma}{N-1}\lti{G}{N-1}+\cdots+\gtti{\Gamma}{0}\lti{G}{0}\nn
&\qquad+\gtti{\Gamma}{N-1}\gti{\Gamma}{N-1}+\cdots+\gtti{\Gamma}{0}\gti{\Gamma}{0}\nn
&=\tti{G}{N-1}\lti{G}{N-1}\nn
&\qquad+\sum^{N-2}_{n=0}\Bigl[\tti{G}{N-1}\lti{J}{N-1}\cdots\lti{J}{N-n}\tti{J}{N-n}\cdots\tti{J}{N-1}\lti{G}{N-1}\Bigr]\nn
&\qquad-\tti{G}{N-1}\Bigl[\gti{\Gamma}{N-1}+\lti{J}{N-1}\gti{\Gamma}{N-2}+\cdots+\lti{J}{N-1}\cdots\lti{J}{1}\gti{\Gamma}{0}\Bigr]\nn
&\qquad-\gtti{\Gamma}{N-1}\Bigl[\lti{F}{N-1}+\gti{\Gamma}{N-1}\Bigr]+\cdots+\gtti{\Gamma}{0}\Bigl[\lti{F}{0}+\gti{\Gamma}{0}\Bigr]\nn
&\qquad+\sum^{N-1}_{n=0}\bigl[\gti{\Gamma}{n}\bigr]^{2}\nn
&=\tti{G}{N-1}M\lti{G}{N-1}-\tti{G}{N-1}\Omega-\sum^{N-1}_{n=0}\bigl[\gtti{\Gamma}{n}\lti{F}{n}\bigr]\label{eq:f2proofmid}
\end{align}
For all $j$ such that $F\ti{n}_{j}\neq 0$, the corresponding $\gamma\ti{n}_{j}=0$ and for all $i$ such that $F\ti{n}_{i}=0$, the corresponding $\gamma\ti{n}_{i}\neq 0$ for all $n$.  Thus $\gti{\Gamma}{n}$ is always orthogonal to $\lti{F}{n}$; therefore $\sum^{N-1}_{n=0}\bigl[\gtti{\Gamma}{n}\lti{F}{n}\bigr]=0$.  Thus, beginning with Eq.~\eqref{eq:f2proofmid} and using Eq.~\eqref{eq:MomegaG},
\begin{align}
F^{2}&=\tti{G}{N-1}M\lti{G}{N-1}-\tti{G}{N-1}\Omega\nn
&=\tti{G}{N-1}\Bigl[M\lti{G}{N-1}-\Omega\Bigr]\nn
&=\lambda\tti{G}{N-1}\lti{G}{N-1}\nn
\therefore \quad R^{2}&=\lambda F^{2}.\label{eq:RlambdaFproof}
\end{align}
Here we have used Eqs.~~\eqref{eq:Rdefn}~and~\eqref{eq:GeqnNbc} to write $R^{2}=\lambda^{2}\tti{G}{N-1}\lti{G}{N-1}$.

\twocolumn


\end{document}